\documentclass[twocolumn, trackchanges]{aastex62}
\hypersetup{linkcolor=red,citecolor=blue,filecolor=cyan,urlcolor=magenta}
\usepackage{xspace}
\usepackage{multirow}

\shorttitle{Spectral-timing of NGC 5506}
\shortauthors{Zoghbi et al.}

\newcommand{\FeK}{\mbox{Fe K$\alpha$}\xspace}

\newcommand{\xmm}{\mbox{XMM-Newton}\xspace}
\newcommand{\nustar}{\mbox{NuSTAR}\xspace}
\newcommand{\rg}{$r_g$\xspace}

\begin{document}

\title{Testing The Lamp-Post and Wind Reverberation Models with XMM-Newton Observations of NGC 5506}

\email{abzoghbi@umich.edu}
\author[0000-0002-0572-9613]{Abderahmen Zoghbi}
\affil{Department of Astronomy, University of Michigan, Ann Arbor, MI 48109, USA}

\author{Sihem Kalli}
\affil{Department of Physics - Mohamed Boudiaf University,  Msila 28000, Algeria}

\author{Jon M. Miller}
\affil{Department of Astronomy, University of Michigan, Ann Arbor, MI 48109, USA}

\author{Misaki Mizumoto}
\affil{Centre for Extragalactic Astronomy, Department of Physics, University of Durham, South Road, Durham, DH1 3LE, UK}
\begin{abstract}
The lamp-post geometry is often used to model X-ray data of accreting black holes. Despite its simple assumptions, it has proven to be powerful in inferring fundamental black hole properties such as the spin. Early results of X-ray reverberations showed support for such a simple picture, though wind-reverberation models have also been shown to explain the observed delays. Here, we analyze new and old XMM-Newton observations of the variable Seyfert-1 galaxy NGC 5506 to test these models. The source shows an emission line feature around 6.7 keV that is delayed relative to harder and softer energy bands. The spectral feature can be modeled with either a weakly relativistic disk line or by scattering in distant material. By modeling both the spectral and timing signatures, we find that the reflection fraction needed to explain the lags is \emph{larger} than observed in the time-averaged spectrum, ruling out both a static lamp-post and simple wind reverberation models.
\end{abstract}


\section{Introduction}\label{sec:intro} 

Observations of several AGN in X-rays have shown signatures of small time delays between the direct and reflected emissions \citep{2009Natur.459..540F, 2010MNRAS.401.2419Z,2013MNRAS.431.2441D,2013MNRAS.430.1408K}. The former is produced in a corona that emits through Compton scattering of lower energy photons \citep{1991ApJ...380L..51H}, while the reflection is produced when the coronal emission illuminates the standard accretion disk \citep{1991MNRAS.249..352G,2014ApJ...782...76G}. 

The magnitude of these lags suggests light-crossing distances of 10-20 gravitational radii ($r_g=GM/c^2$) at most, implying a very compact corona \citep[e.g.][]{2012MNRAS.422..129Z, 2016MNRAS.462..511K}. The delays of the \FeK line \citep{2012MNRAS.422..129Z,2014ApJ...789...56Z,2014MNRAS.438.2980C,2016MNRAS.462..511K}, which is in a relatively clean part of the spectrum compared to the soft band ($<$ 1 keV), are of a particular importance. The lag measurements in most cases are simple, amounting to a single number, representing the average delay between the direct and reflected emissions at some variability time scale.\citep[see ][for a review]{2014A&ARv..22...72U}. 

Significant progress has been made in attempting to model observed delays with a point corona that illuminates a thin disk (i.e. lamp-post) \citep{2014MNRAS.438.2980C, 2014MNRAS.439.3931E}, and an extended one \citep{2016MNRAS.458..200W}. Several studies have also attempted modeling the delays simultaneously with the variable spectrum (the RMS or the covariance spectra; e.g. \citealt{2014A&ARv..22...72U}) and the total time-averaged spectrum \citep{2016MNRAS.460.3076C,2018MNRAS.475.4027M,2019MNRAS.488..324I}. Such modeling has however been challenging given the weak reverberation signals measured in most cases, and the complexity of the models.

Attributing the observed delays uniquely to relativistic reverberation, albeit simple and attractive, is not always possible, particularly in the soft band ($<1$ keV) where a direct association of the delays with atomic features is not trivial. For instance, \cite{2010MNRAS.408.1928M} attributed the delays between the soft band ($<1$ keV) and the continuum-dominated band ($1-3$ keV) to reverberation from a large-scale system of reprocessing clouds close to and off the line of sight. The small delays are caused both by the presence of significant reprocessing material close to the line of sight, and by an artifact of the Fourier-based delay measurements. The delays in the soft band  have also been explained by models that attributes the soft excess seen in many sources to a warm corona (in addition to the standard hard corona) instead of it being dominated by reflection \citep{2012MNRAS.420.1848D}.

A key prediction of the relativistic reverberation model in the \FeK band  is that the shape of the lag profile with energy should match the shape of a broadened iron line. The number of sources where this comparison was made has been limited (e.g. \citealt{2012MNRAS.422..129Z,2014MNRAS.438.2980C}, but see also \citealt{2019AZ_4151} for more recent data). It is particularly known (and also expected) that not all AGN have a very strongly broadened \FeK lines \citep{2007MNRAS.382..194N, 2012MNRAS.426.2522P}. This leads to a simple hypothesis: If the observed lags are due to relativistic reverberation, then the shape of the lag-energy spectra should match the shape of the line profile in the time-averaged spectrum. A relatively narrow relativistic \FeK line in the time-average spectrum should correspond to longer delays and a narrow lag-energy profile. In 2015, the nucleus of NGC 5506 was targeted with an \xmm observation to test this hypothesis. This work reports the analysis of that and previous observations.

NGC 5506 is a nearby ($z=0.00618$) X-ray obscured ($N_h=3\times10^{22}\rm{cm}^{-2}$; \citealt{1999ApJ...515..567W} Narrow Line Seyfert 1 galaxy \citep{2002A&A...391L..21N}. At soft energies (below 1 keV), the spectrum is dominated by scattering and reprocessing of the nuclear radiation by photoionized gas originating on scales as large as a few hundred parsecs \citep{2003A&A...402..141B}. Early analysis of XMM-Newton and Chandra observations showed no evidence of a relativistic iron line from the accretion disk \citep{2003A&A...402..141B}, but other analyses \citep{2001A&A...377L..31M,2007MNRAS.382..194N} attributed the absence to the low signal to noise ratio in the spectra. An analysis of longer observations showed the presence of a broad component of the iron line \citep{2010MNRAS.406.2013G,2018MNRAS.478.1900S}. The line was not very broad, indicating a slowly- or non-spinning black hole. This was also the conclusion reached earlier by a long ASCA observation \citep{1999ApJ...515..567W}.

\section{Observations \& Data Reduction}\label{sec:data} 

\begin{deluxetable}{ccllll}
\tablecaption{Description of the observed data.\label{tab:obs_log}}
\tablehead{\colhead{\#} & \colhead{ObsID} & \colhead{Exp. (ks)} & \colhead{Time (MJD)}}
\startdata
1       & 0013140101 & 13.5	    & 51943.0 \\
2       & 0013140201 & 9.9	    & 52283.8 \\
3       & 0013140201 & 14.8	    & 53197.5 \\
4       & 0201830301 & 14.0	    & 53201.0 \\
5       & 0201830401 & 13.9	    & 53208.7 \\
6       & 0201830501 & 14.0	    & 53225.0 \\
7       & 0554170101 & 57.7	    & 54834.3 \\
8       & 0554170201 & 62.9	    & 54674.8 \\
9       & 0761220101 & 88.8	    & 57211.7 \\
\enddata
\end{deluxetable}

NGC 5506 has been observed several times with \xmm. The most recent observation was in July 07, 2015. We reduced this and all prior observations (a total of 9 exposures). The \xmm EPIC data were reduced using \texttt{epchain} in \textsc{SAS}. Multiple exposures within a single observations are combined. 

Source and background photons are extracted from circular regions of 50\arcsec\,radius centered on and away from the location of source respectively. Given the brightness of the source, photon pileup needs to be considered. We use the \texttt{epatplot} tool to check for the effect. We found that the ratio of observed to predicted single and double photon events are consistent with unity, implying no strong pileup. All EPIC spectra are grouped so that the detector resolution is over-sampled by a factor of 3, ensuring the minimum signal to noise ratio per bin is 6.
Details of the \xmm observations are shown in Table \ref{tab:obs_log}. For the spectral modeling presented in the following sections, we use the {\sc xspec} for the spectral modeling, and employ $\chi^2$ statistics for model fitting.

\begin{figure}
\includegraphics[width=\columnwidth]{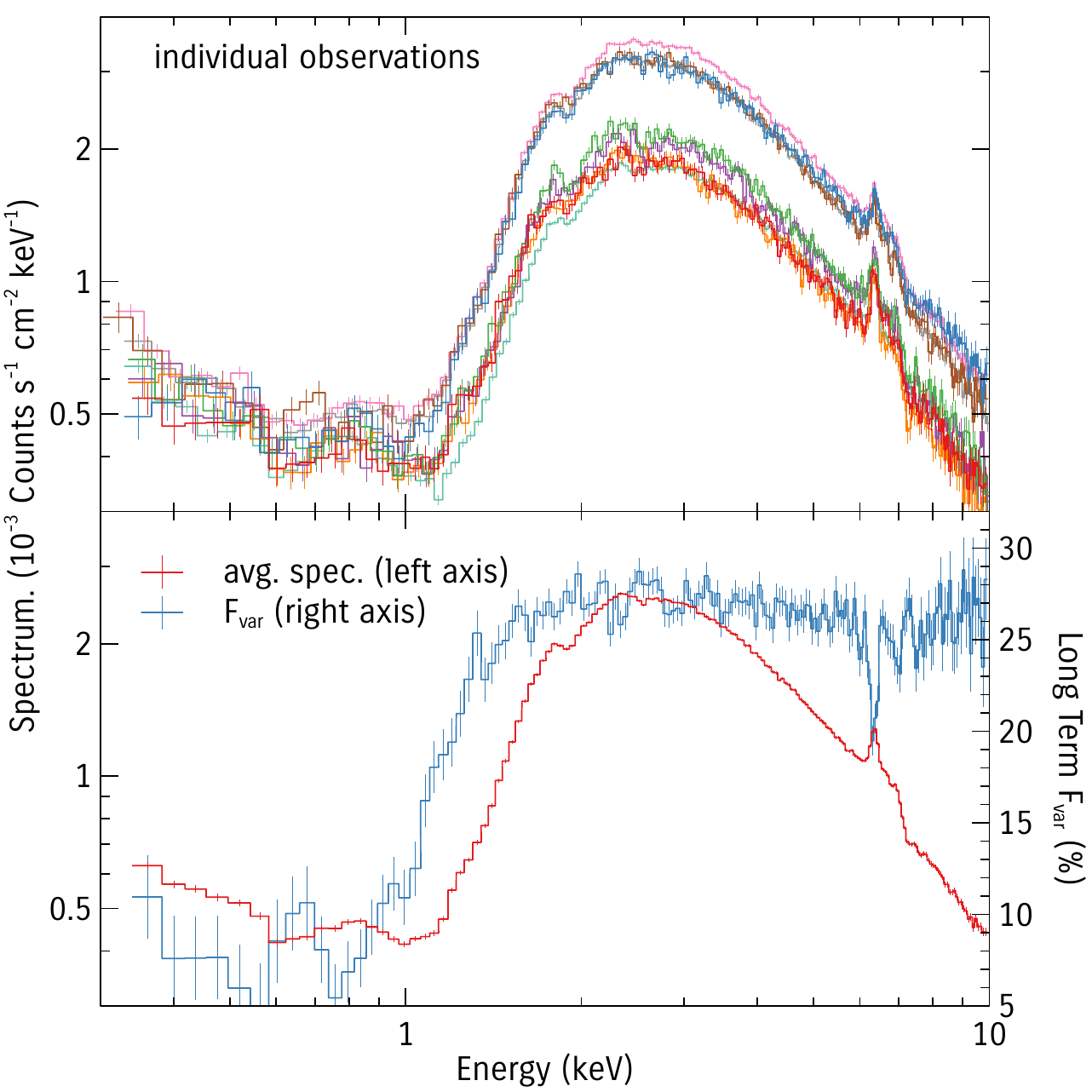}
\caption{{\em Top:} The spectra from all 9 observations plotted after factoring out the detector area. {\em Bottom:} The average spectrum from all observations along with the long term (i.e. between observations) fractional variability amplitude $F_{\rm var}.$\label{fig:gen_shape}}
\end{figure}

For the timing analysis, we only use observations with the difference between start and stop times is at least 20 ks, which excludes the first two. In order to obtain the longest and most continuous light curves possible, we select only good time intervals where the background rate between 10--12 keV is below 0.5 counts per second. This slightly relaxed from the 0.4 counts per second in the standard spectral extraction. Light curves in the energies of interest are extracted by filtering on the PI values, then we use \texttt{epiclccorr} to apply both absolute (vignetting, bad pixels, chip gaps, PSF and quantum efficiency) and relative (deadtime, GTI, exposure and background) corrections. All the light curves analyzed in this work are background-subtracted using the same extraction regions from the spectral analysis. 

\section{Spectral Analysis} \label{sec:spec}

\subsection{The General Shape}\label{sec:spec:general}
The long term variations in the spectrum of NGC 5506 are characterized by a roughly constant shape that varies in flux. This is illustrated in Figure \ref{fig:gen_shape}, which shows the spectra from all the 9 EPIC-PN exposures, the effective area curve has been factored out by unfolding the spectra to a constant. 
The plot shows that the shape is roughly constant over the years while the flux changes by $\sim30\%$ above 2 keV.
As pointed out by \cite{2010MNRAS.406.2013G}, the absorption does not appear to change with time as indicated by the same factor change at 2 and 10 keV, suggesting that the absorption is distant and likely produced in the large dust lane seen across the optical image of the galaxy \citep{2011A&A...532A..74B}. The soft part of the spectrum ($< 2$ keV), which is dominated by photoionized gas emission originating at scales of hundred parsecs from the nucleus, shows very little variability. The lower panel of Figure \ref{fig:gen_shape} shows the long term fractional variability amplitude \citep[e.g.][]{2003MNRAS.345.1271V} measured between observations. It has a constant shape between 2--10 keV implying a constant spectral shape. It drops below 1 keV because the spectrum is not variable in that band. It also shows a dip at 6.4 keV, corresponding to a less variable narrow \FeK line, with less clear structure between 6--7 keV.

\begin{figure}
\includegraphics[width=\columnwidth]{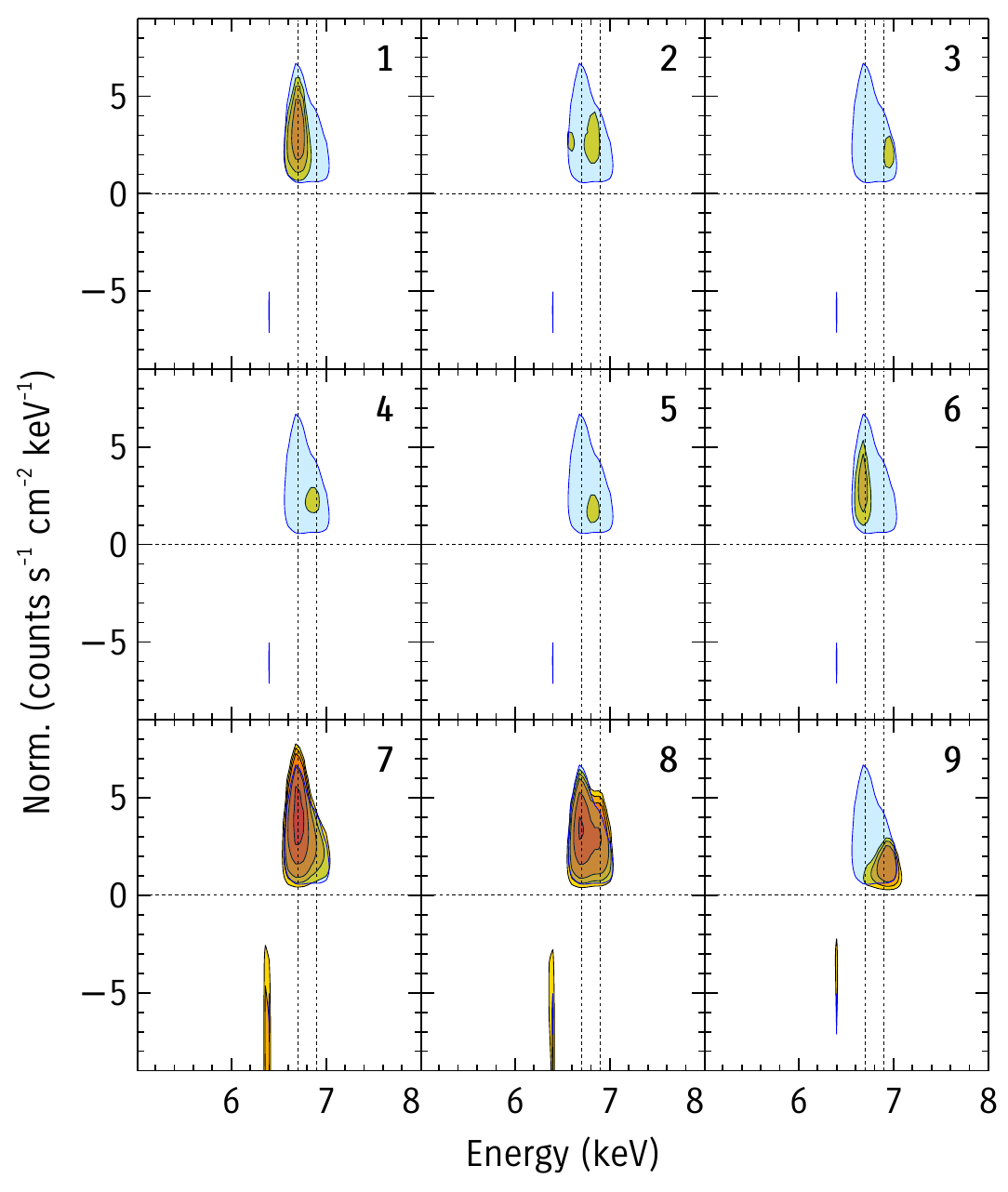}
\caption{The residuals to the basic fit shown as contours of feature significance as a function of energy and the intensity of narrow Gaussian function at each point. The significance contours are shown in units of $\sigma$, showing the levels 2, 3, 4, 6, 8, 10 and 20, i.e. a value of 3 means the feature at that energy and intensity is significant in the residuals at $3\sigma$. The blue shade is for the combined residuals showing the $3\sigma$ contour line, while the red-yellow shade is for individual observations as labeled in each panel. The two vertical dotted lines in each panel are at 6.7 and 6.9 keV for a guide.\label{fig:line_scan}}
\end{figure}

\subsection{Spectral Modeling}\label{sec:spec:mod}
We focus in the spectral modeling on the nuclear emission dominating the 2--10 keV band. The spectrum below 2 keV is not directly emitted in the nuclear region \citep{2003A&A...402..141B}. A basic model would include an absorbed power law to model the primary coronal emission (modeled with {\tt ztbabs*powerlaw}), and a model for the strong narrow \FeK emission line. The latter component can originate in Compton-thin material such as the broad or narrow line regions (BLR or NLR), or Compton-thick material like the outermost regions of the disk or a torus. The two can be distinguished by the Compton Reflection (CR). \cite{2001A&A...377L..31M} used the edge of the CR to imply that the line-emitting region is compton-thick with optical depth larger than unity. We model this line using the {\tt xillver} model \citep{2014ApJ...782...76G}. The starting base model therefore has the form: {\tt tbabs*(ztbabs*powerlaw + xillver)}, where the {\tt tbabs} model accounts for the Galactic absorption column density, fixed at $4.1\times10^{20}\ {\rm cm}^{-2}$. Here, we make the inconsequential assumption that the reflection is outside the local absorber. Including the reflection inside the absorber only changes the inferred intrinsic flux of the narrow \FeK line. Fitting the nine spectra separately, we find that this model accounts for the general shape of the spectrum, but leaving strong residuals peaking at 6.7 keV.

Given the detector resolution, it is not clear whether these residuals are due to a single broadened ionized disk emission, or a blend of narrow emission lines (H-like and He-like iron). These latter lines can be produced by fluorescence and resonant scattering in photoionized matter \citep{1996MNRAS.280..823M}, and likely produced in the region producing the spectrum below 1 keV.

\begin{figure*}
\includegraphics[width=\textwidth]{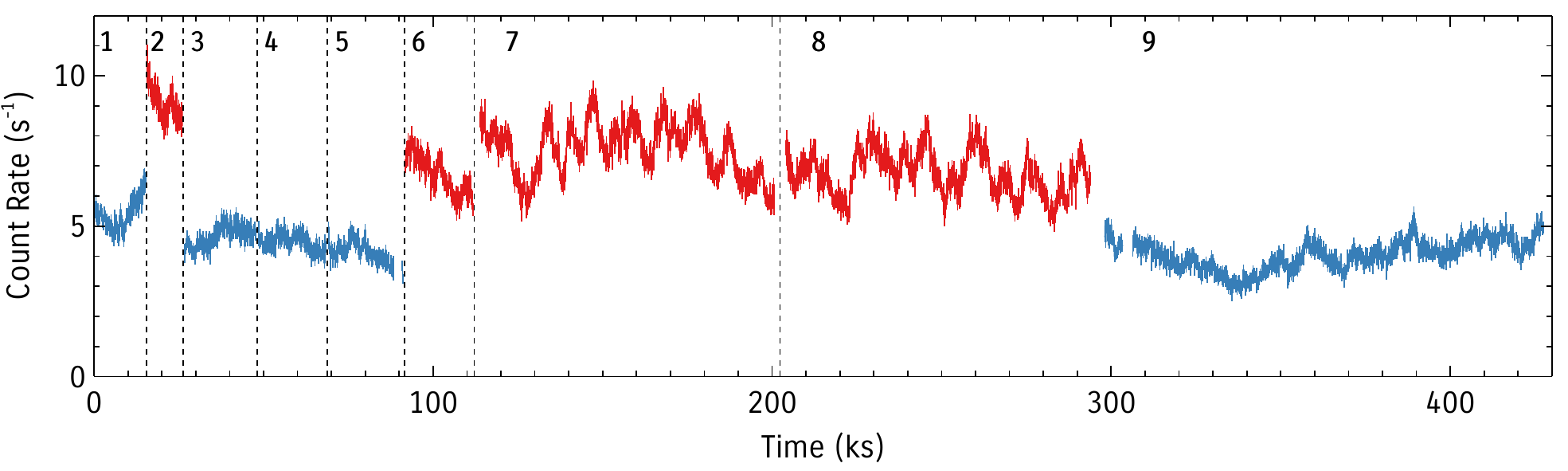}
\caption{Light curves from all observations. Blue and red colors indicate the low and high flux intervals. \label{fig:lc} }
\end{figure*}

To quantify the significance of the residuals in each observation, we show them in Figure \ref{fig:line_scan} as contours of significance levels in the energy-intensity space, produced by adding a narrow Gaussian function to the base model with a grid of energy and intensity values. The significance was calculated following the prescription in \cite{2015ApJ...799L..24Z}, where we find the $\Delta\chi^2$ improvement at grid point, we then fake 10,000 spectra assuming model parameters from the base model and their uncertainties, and find the distribution of $\Delta\chi^2$ that are due to the observational statistics. These $\Delta\chi^2$ values are considered at {\it any} observed energy, so the number of trials is properly accounted for. The observed $\Delta\chi^2$ is then compared to this simulation distribution to obtain a significance. The nine spectra are assessed independently.

Figure \ref{fig:line_scan} shows that the 6.7 keV feature is detected at more than $3\sigma$ confidence in 5 out of 9 observations. The weaker detection in the other four is likely because of their low exposure and low count rate. The apparent absorption line at 6.4 keV in the last three observations is an artifact of the base model trying to model the strong unmodeled residuals at 6.7 keV.



The residuals can be modeled as a sum of two narrow unresolved Gaussian functions at 6.7 keV and 6.9 keV (i.e. their width is consistent with the detector resolution) or a single broad line. The narrow lines can be due to recombination lines from Fe{\sc-xxv} and Fe{\sc-xxvi} respectively \citep{2010MNRAS.406.2013G}.
When fitted with two narrow lines, the 6.9 keV line is consistent with having a constant flux, while the constant flux hypothesis for the 6.7 keV line is ruled out at more than 99.99\% confidence (This can also be seen from Figure \ref{fig:line_scan} by comparing observations 8 and 9 for instance).

There are indications in the data that the total flux from these residuals (the sum of two narrow lines or a single broad line) is correlated with the total continuum flux. The Spearman's rank correlation coefficients are 0.57 and 0.65 when using the fluxes from the narrow lines and the broad line compared to the 7--10 keV flux, with the no-correlation hypothesis rejected at the 95 and 98\% confidence, respectively.

Distinguishing between the narrow lines vs a broad line models is not possible based on spectroscopy alone, and this is the origin of the discrepancy in interpreting the spectra of NGC 5506 in the literature \citep[e.g.][]{2003A&A...402..141B,2010MNRAS.406.2013G,2018MNRAS.478.1900S}. The suggestion that the line flux is correlated with the continuum flux may be an indication that the broad line origin is more likely. In section \ref{sec:timing}, we include additional information from the fast variability. For the completeness of the spectral analysis, we also model the spectra using a full relativistic model and present the model parameters next.

\subsubsection{Relativistic Model} \label{sec:rel_model}
We model the relativistic reflection using {\tt relxill}, which is a combination of a reflection code {\tt xillver} \citep{2010ApJ...718..695G} and the {\tt relline} ray tracing code \citep{2010MNRAS.409.1534D}. All the spectra are modeled simultaneously. The model has the {\sc xspec} form: {\tt tbabs*(ztbabs*(relxill+powerlaw) + xillver)}, with the other components similar to those discussed at the start of section \ref{sec:spec:mod}. We start by assuming that only the flux and photon index of the primary power law, and the flux of the relativistic reflection change in time, in addition to the line of sight absorption. Because the Fe emission line is not very broad, it cannot constrain the spin, so we fix it at maximum\footnote{Fixing it at 0 does not change the results because, as we will show, the inner radius of emission is relatively large.} and fit for the inner radius of the accretion disk instead. We assume a single emissivity index that covers the whole disk extending from the inner radius (a free parameter) to 1000 gravitational radii ($r_g=GM/c^2$). We find through initial modeling that allowing non-solar abundances both in the reflection and the absorption (assumed to be the same) provide a significant improvement so it is allowed to vary during the fit. Additionally, and to obtain a better handle on the spectrum above 10 keV, we include data from one publicly available \nustar observation (obsID 60061323002). 

\begin{figure}
\centering\includegraphics[width=\columnwidth]{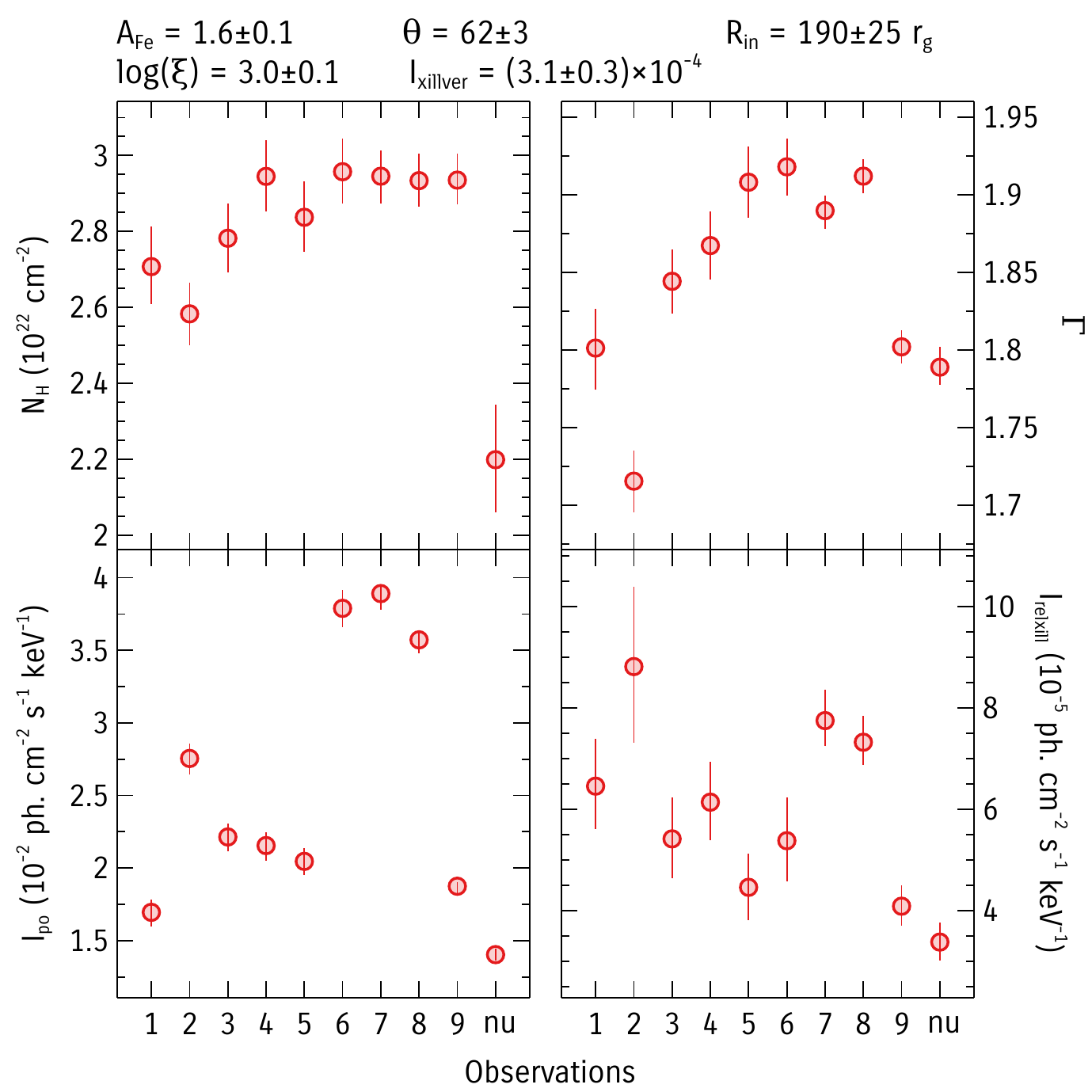}
\caption{Parameters of the relativistic model.\label{fig:rel_fit}. Parameters at the top of the panels are those assumed to be non-variable between observations ($A_{\rm Fe}$ is the iron abundance, $\theta$ is the inclination of the inner disk, $R_{\rm in}$ is the inner extent of the disk, $\xi$ is the ionization parameter and $I$ refers to the normalization). The panels shows the variable parameters ($N_{\rm H}$ is the absorption column density, $\Gamma$ is the photon spectral index, $I_{\rm po}$ and $I_{\rm relxill}$ are the normalizations).}
\end{figure}

The parameters of the best fitting model are summarized in Figure \ref{fig:rel_fit}, where the non-variable parameters are shown at the top, while the variable parameters are plotted in the four panels. The plots show the values for the 9 XMM-PN exposures plus one \nustar (the FPMA/FPMB were fitted simultaneously assuming all parameters are the same except for a cross calibration constant, whose best value is found to be $1.00\pm0.01$). The best fit statistic is $\chi^2 = 2919$ for 2678 degrees of freedom, corresponding to a reduced $\chi^2_\nu$ of 1.09.

As can be seen from Figure \ref{fig:rel_fit}, the intensity of the primary power law component is the most variable parameter followed by the intensity of the 6.7 keV feature modeled by {\tt relxill}. $\Gamma$ changes so that the spectra are slightly softer at higher fluxes, as commonly observed in other Seyfert galaxies \citep[e.g.][]{2004A&A...422...85P, 2009MNRAS.399.1597S}. The absorption column density shows much less variability, and it is mostly constant around $3\times10^{22}$ cm$^{-2}$ except in the \nustar observation, where it drops to $2.2\times10^{22}$ cm$^{-2}$. Given the energy resolution and coverage of \nustar which extends down to 3 keV only, the $N_{\rm H}$ value inferred is more uncertain. If we fix the column density in the \nustar spectra to the average from all nine \xmm observations, the fit statistic increases by $\Delta\chi^2=-19$ for 1 additional degree of freedom. No significant change is observed in the other parameters.

The best fit indicates that the inner disk is seen at an intermediate inclination of $\theta=62\pm3^{\circ}$ and has a slightly above solar iron abundance. The inner radius of the disk is at $R_{\rm in}=190\pm25 r_g$. Allowing the inner disk to vary between observations improved the fit by $\Delta\chi^2 = 21$ for 9 additional free parameters. The null hypothesis probability for a non-variable $R_{\rm in}$ using the f-test is  $p=0.023$, indicating that it is rejected at 97.7\% confidence. We also tested for the alternative hypothesis that $R_{\rm in}$ varies only between two flux states (the cut is at power law intensity of $I_{\rm po} = 2.5\times10^{-2}$, see Figure \ref{fig:rel_fit}). The two best fit values are $134\pm22$ and $241\pm43 r_g$ for the low and high flux observations, respectively. The improvement in $\chi^2$ relative to the constant $R_{\rm in}$ model is $\Delta\chi^2 = 6$ for one additional free parameter, corresponding to a rejection probability for the non-variable hypothesis of 98.1\% confidence. We consider these as suggestive evidence for changes in $R_{\rm in}$. Allowing the inclination angle to change between the low and high flux intervals improves the fit by $\Delta\chi^2=3$ for one degree of freedom, which corresponds a null hypothesis probability for a non-variable inclination of 0.097, which is weaker than the $R_{\rm in}$ changes.

If we fix the reflection fraction to that expected from a lamp-post geometry, we can fit directly for the height of the corona using the {\tt relxilllp} model. We find a disk inner radius of $R_{\rm in} = 178\pm25 r_g$, and height of $h=95^{+61}_{-7} r_g$, with the rest of the parameters consistent with those in Figure \ref{fig:rel_fit}. Allowing for $R_{\rm in}$ and $h$ to vary with time or with flux did not provide a significant improvement to the fit. {\it This analysis implies that the spectral data is consistent with the static lamp-post model.}

\section{Timing Analysis} \label{sec:timing}
NGC 5506 is known for its strong variability \citep{1987Natur.325..696M}. The 2--10 keV light curve from all nine \xmm observations are shown in Figure \ref{fig:lc}. The high and low flux intervals identified from the spectral modeling in section \ref{sec:spec} are plotted in different colors.

\subsection{Power Spectrum}\label{sec:timing:psd}
We start by estimating the power spectrum density (PSD) using the periodogram following the standard procedures \citep[e.g.][]{2003MNRAS.345.1271V}. 
Using all the observations, we find that the periodogram strongly favors models that include a break around 0.1 mHz. Fitting a bending power law model with the lower index fixed at 1 gives a break frequency of $0.19\pm0.04$ mHz, and a high frequency slope of $a_2 = 3.2\pm0.4$, both consistent with published work \citep[][using observations 7 and 8 only]{2012A&A...544A..80G}. Fitting the same model to the high and low flux intervals separately gives parameters that are consistent with the those found using all the data, therefore, significant non-stationary behavior is ruled out.

\subsection{Covariance Spectra}\label{sec:timing:cov}
We also calculate the covariance spectra, which measure the shape of the correlated variability at some Fourier freuquency \citep[e.g.][]{2014A&ARv..22...72U}. We use two frequency bins $0.008-0.1$ and $0.1-0.26$ mHz and refer to them as the low and high frequency bins. The justification of the bin choice is based on the lag measurements discussed in section \ref{sec:timing:delays}. We find that the covariance spectra can be described by an absorbed power law model for both frequency bins. The fit parameters provide evidence that the continuum is significantly harder at higher frequencies: $\Gamma=1.91\pm0.03$ vs $1.75\pm0.05$ for the low and high frequency bins respectively. This is consistent with observations of other sources that show flatter PSDs at higher energies \citep{2007MNRAS.382..985M, 2011MNRAS.412...59Z,2012A&A...544A..80G}, and it is an indication that the variability is driven by the primary continuum, not by other processes such as intervening absorption. 

Adding a Gaussian line to the covariance spectra around 6.7 keV did not provide significant fit improvement. This applies to the whole dataset and also when considering the high and low flux data separately.
The $90\%$ confidence upper limits on the reflection fraction\footnote{In this work, the reflection fraction refers to the observed ratio between the reflected emission and the continuum, so the it can be used in the context of both relativistic reflection and wind scattering. See \cite{2014MNRAS.444L.100D} for a useful discussion.} at the line peak (assuming the energy and width are similar to the time-averaged spectra) are: $(6\%, 14\%)$ for the low and high frequency bins respectively. 
For comparison, the fraction of the line contribution, also at the peak, from the time-averaged spectrum is $5.4\pm0.4\%$.
The main conclusion here is that the covariance spectrum provides only upper limits on the fractional variability of the 6.7 keV feature, and these limits are higher than its fractional flux contribution to the time-averaged spectrum. 
Assuming that the line varies only as a response to continuum variability (i.e. it cannot vary more than the continuum), the reflection fraction in the time-averaged spectrum itself can be used as an upper limit on the variable reflection fraction.


\begin{figure}
\centering\includegraphics[width=0.8\columnwidth]{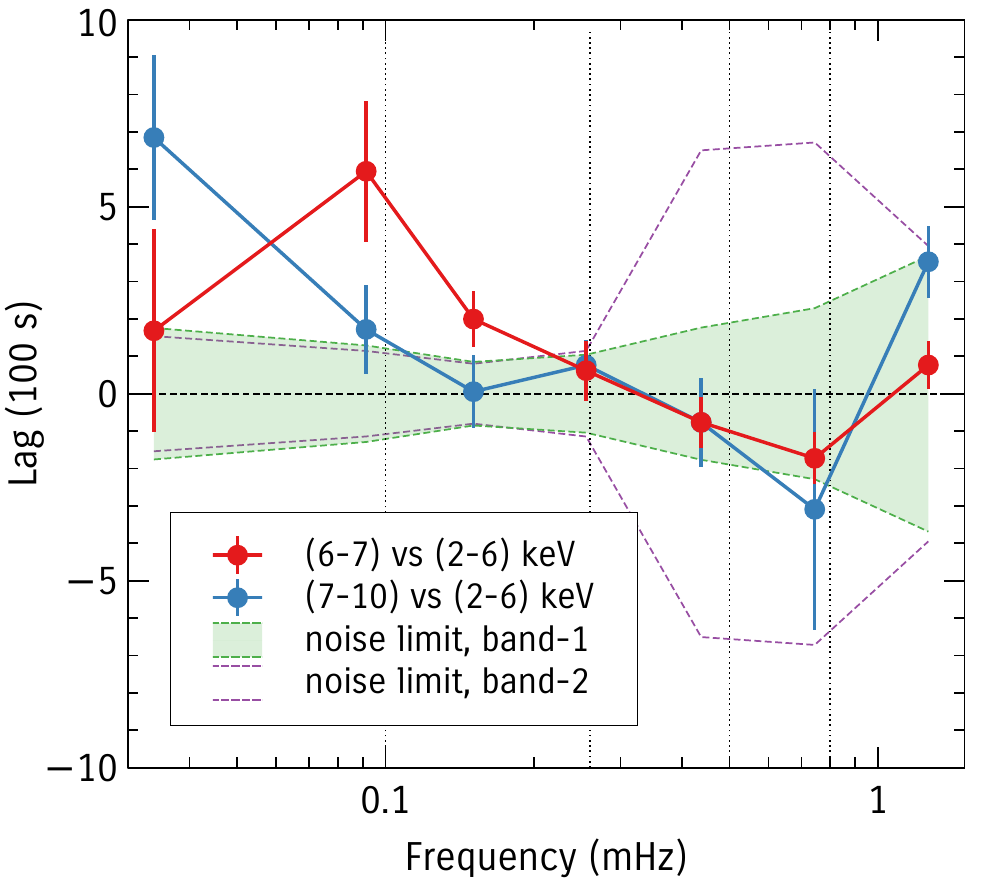}
\caption{Lag vs frequency from all the data comparing the iron band (6--7 keV) and the hard band (7--10 keV) with the 2--6 keV band. Positive lags indicate the hard band lags the softer band. 
The $1\sigma$ limits on the lag measurements under the assumption of a zero delay \citep[equation 30 in][]{2003MNRAS.339.1237V} are also shown.\label{fig:lag_fq}}
\end{figure}

\subsection{Time Delays}\label{sec:timing:delays}
Time lags generally depend on both frequency and energy. For the frequency dependence, Light curves at three energy bands, 2--6, 6--7 and 7--10 keV, are extracted. The first and the last bins are generally dominated by the continuum, and the second bin is dominated here by the emission from the 6.7 keV feature. The light curves are tapered with a Hanning function \citep{Bendat:2000:RDA:555747} to reduce red noise leak effects before calculating time lags following the standard procedure \citep{2014A&ARv..22...72U}. The frequency axis was binned with a geometric factor of 1.7, requiring that each frequency bin contains at least 15 independent Fourier frequencies. The result is shown in Figure \ref{fig:lag_fq}. Positive lags indicate the hard band lags the softer band. Figure \ref{fig:lag_fq} shows that below $\sim0.3$ mHz, where the signal is not dominated by Poisson noise, the iron band, 6--7 keV, appears to be delayed with respect to the 2--6 keV band between 0.1--0.2 mHz, while the 7--10 keV is the one that lags the 2--6 keV at lower frequencies.

\begin{figure}
\centering\includegraphics[width=\columnwidth]{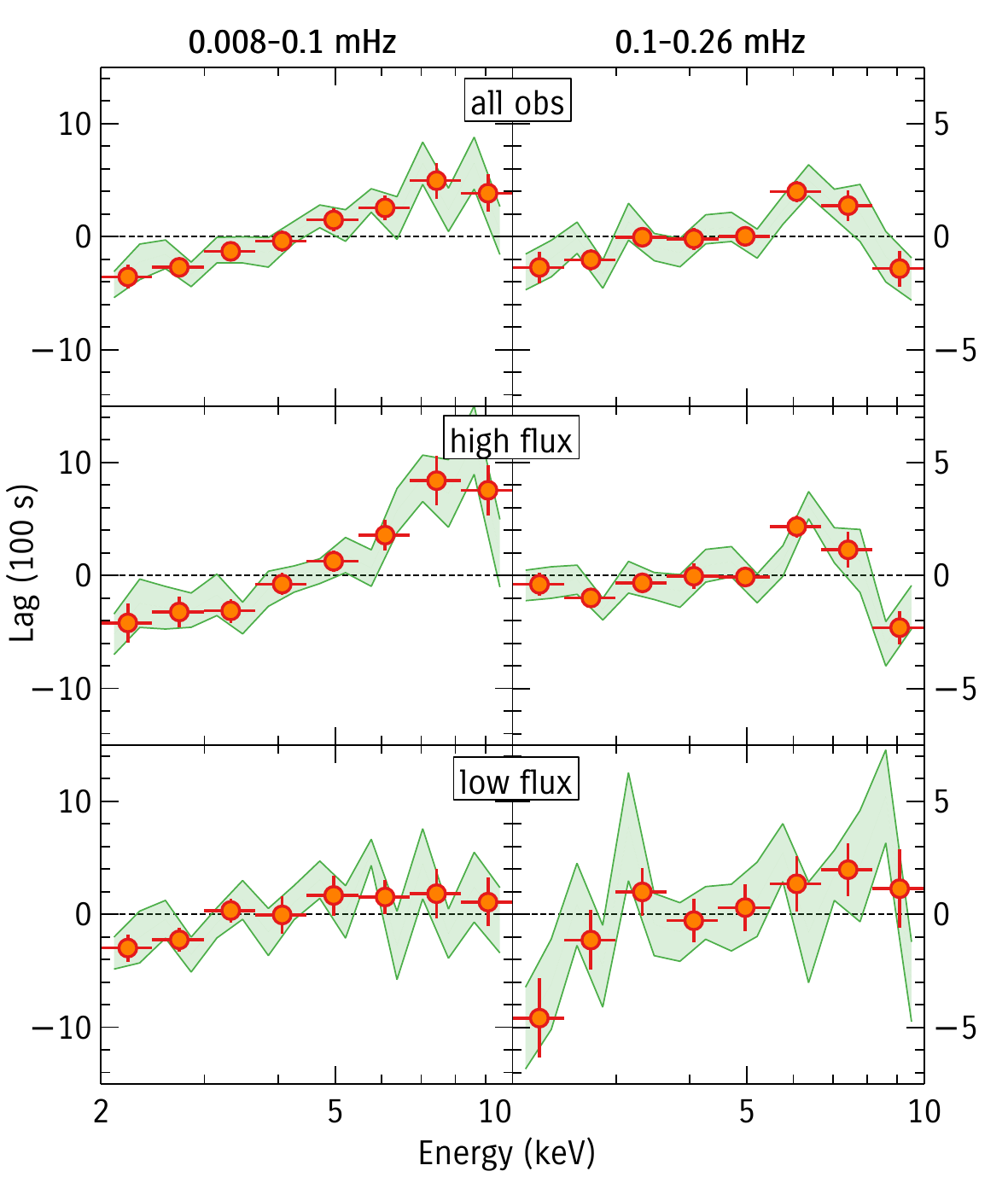}
\caption{Lag vs energy at two frequency bins for the total, high and low flux intervals. Note the difference in the y-axis between the left and right columns.\label{fig:lag_en}}
\end{figure}

Given these indications, we calculate the lag-energy dependence at two frequency bins $0.008-0.1$, and $0.1-0.26$ mHz. The orange points in Figure \ref{fig:lag_en} show the results of calculating the lags using 8 energy bins between 2 and 10 keV in log-steps. The top panels show the result when using all observations while the middle and lower panels are for the high and low flux intervals respectively. The shaded green regions show the lags produced when using 16 energy bins, and explores the effect of the energy binning choice.

The plots in Figure \ref{fig:lag_en} show that there are clear deviations from zero lag. The significance of these measurements against a constant and a linear (vs log-energy) null models are shown in Table \ref{tab:lag_tests}. The constant model tests for the presence of {\em any} inter-band delays. The log-linear model tests the data against a feature-less trend. 

Table \ref{tab:lag_tests} shows that the constant model is rejected at more than $99\%$ confidence in all cases. The log-linear model is only rejected in the high frequency bin of the combined dataset and the high flux subset. The feature-less lag at the low frequency bin appears to be similar to the continuum lags that often dominate at low frequencies in many sources \citep{1989Natur.342..773M,2001MNRAS.327..799K,2006MNRAS.367..801A,
2011MNRAS.412...59Z}.

The only case therefore where complexity beyond the two simple models is required is the high frequency bin of the high flux observations (in addition to the combined observations, whose lags are driven by the high flux data). Modeling the additional complexity with a Gaussian function gives a centroid energy of $E = 6.5\pm0.1$ keV. This energy suggests that these lags are related to the 6.7 keV spectral feature found in the time-averaged spectra discussed in sections \ref{sec:spec:mod}. 

\begin{deluxetable}{l|ll|ll}
\tablecaption{Results of the hypothesis test for the lag spectra in Figure \ref{fig:lag_en}. The table shows the null hypothesis $p$ value from each test for a \texttt{constant} and a \texttt{linear} models. Values in bold are those below $p=0.01$, corresponding to a 99\% confidence in rejecting the null hypothesis. The tests are done using the 16 bins data. Other binning schemes give comparable results. \label{tab:lag_tests}}

\tablehead{\multirow{2}{*}{Obs} & 
\multicolumn{2}{c|}{\texttt{constant}} & 
\multicolumn{2}{c}{\texttt{linear}} \\
& bin-1 & bin-2 & bin-1 & bin-2}
\startdata
all	obs.	& $\bf{3(10^{-8})}$  & $\bf{2(10^{-4})}$ & $0.40$         & $\bf{4(10^{-3})}$ \\
high flux	& $\bf{7(10^{-10})}$ & $\bf{1(10^{-6})}$ & $0.16$         & $\bf{5(10^{-6})}$ \\
low flux	& $\bf{1(10^{-4})}$  & $\bf{6(10^{-3})}$ & $1.8(10^{-2})$ & $5.7(10^{-2})$ \\
\enddata
\end{deluxetable}

\subsection{Lag Interpretation}\label{sec:timing:lag_interp}

Interpreting the delay values requires that the observed inter-{\em band} lags are converted to {\em component} delays by accounting for their relative contribution to the variable spectrum at different energies (i.e. the reflection fraction). Before using a lamp-post model to fit the lag data, we consider {\it a model-independent} dilution correction first.

Based on the discussion in section \ref{sec:timing:cov}, we use the upper limit on the reflection fraction from the time-averaged spectrum. We use the best fitting models for the time-averaged spectra to calculate the $99.7\%$ confidence upper limit on the fractional contribution of the reflection component at each energy band $f_r(E)$. 
Two cases are considered for modeling the reflection line: a Gaussian function, and a full {\tt relxill} model (see section \ref{sec:spec:mod}). Then a model of the form $a + b\times f_r(E)$ is used to fit the lag spectra, where $a$ and $b$ are model parameters. $a$ is a shift parameter that accounts for the reference band in the lag calculations, and $b$ measures the delay between the reflection and direct components. This modeling assumes there are no inter-band delays within the continuum and the reflection spectra themselves.

Fitting this model to the high-frequency high-flux lag data, we obtain lower limits ($99.7\%$ confidence) on the dilution-corrected lag of $b>2755$ seconds when using a Gaussian function, and $b>2106$ seconds when using the full {\tt relxill} model. The limit is smaller when using the full reflection model because reflection contributes to the whole reference band, not just where the line is present, and that dilutes the delays. Because we are discussing lower limits on the lag (given the upper limits on the reflection fraction), we consider the smaller value from the {\tt relxill} model, giving a $99.7\%$ confidence lower limit on the dilution-corrected lag of $\tau > 2106$ seconds.

Now we consider the frequency at which the lags are measured. A time lag $\tau$ cannot be measured at frequencies higher than the wrap frequency $\nu_w = 1/2\tau$, because at $\nu_w$, a lag of $\tau$ is the same as a lead of $-\tau$, as the phase delay $\phi=2\pi\nu_w\tau=\pi$, and a phase lag of $\pi$ is the same as a lead of $-\pi$. In fact, this is the case for all frequencies $\nu_n = n/2\tau$ where $n\geq1$ is an integer. Also, $\tau$ here is the dilution-corrected value, not the observed inter-band lag.

Given the lower limit on the dilution-corrected lag of $\tau > 2106$ seconds, the first ($n=1$) wrap frequency is $\nu_w < 0.24$ mHz. This is comparable to the frequencies in which the lag is observed ($0.1-0.26$ mHz). implying that direct interpretation of the lags requires careful considerations.

First, we consider the possibility that the lag we measure is {\em above} the wrap frequency. This would suggest that the observed lag is in the oscillatory part of the delay transfer function, and the actual intrinsic lag is much larger than the observed lags. This would also mean that the lag of the iron band relative to the continuum instead of a lead is just a coincidence. A prediction of this possibility is that the lag should flip sign when we probe lower frequencies, the iron band will be leading instead of lagging due to the oscillatory nature of the transfer function.

This prediction can be tested given that we have data in one lower frequency bin, namely 0.008--0.1 mHz, so we can test for the presence of the 6.7 keV feature and for the change of the lag sign. Unlike the high frequency lag data, the low frequency has a trend that increases with energy. Accounting for the trend, we find that the presence a feature at 6.7 keV in the low frequency band that is similar to the one at high frequencies cannot be ruled out. The contribution in the low frequency bin is constrained to be a factor $f=0.8\pm0.5$ of the lag at the high frequency bin. In other words, we do not see any significant change in the lag of the 6.7 keV feature in the low frequency bin, and a flip in the sign ($f=-1$) can be ruled out at more than $99.9\%$ confidence.

This implies that the lag of the 6.7 keV feature is present at least down to 0.008 mHz, or a factor of $\sim30$ in frequency (see also Figure \ref{fig:lag_fq}), implying that we are {\em not} observing the lag in the oscillatory part of the transfer function, at least for simple reverberation models \citep[see similar discussion in][]{2011MNRAS.412...59Z}, because the lag is expected to switch sign at multiples of $\nu_w$, and the smaller its value, the more sign changes should be observed. The question of whether there is a sign flip at frequencies lower than 0.008 mH remains open, and future observations that target time scales $>100$ ks will be able to address it.

\subsection{Lag Modeling}\label{sec:timing:lag_model}
In this section we attempt to model the iron band lag, taking the information from the time-averaged spectrum into consideration. In section \ref{sec:rel_model}, we showed that the 6.7 feature in the time-averaged spectrum can be modeled with weakly relativistic disk reflection in a lamp-post geometry. We use the reverberation model {\tt kynxilrev}\footnote{Available from \url{https://projects.asu.cas.cz/stronggravity/kynreverb}}, which  computes the time dependent reflection spectra of the disc as a response to a flash of primary power-law radiation from a point source located on the axis of the black-hole accretion disc \citep{2004ApJS..153..205D}.

\begin{figure*}
\centering\includegraphics[width=0.8\textwidth]{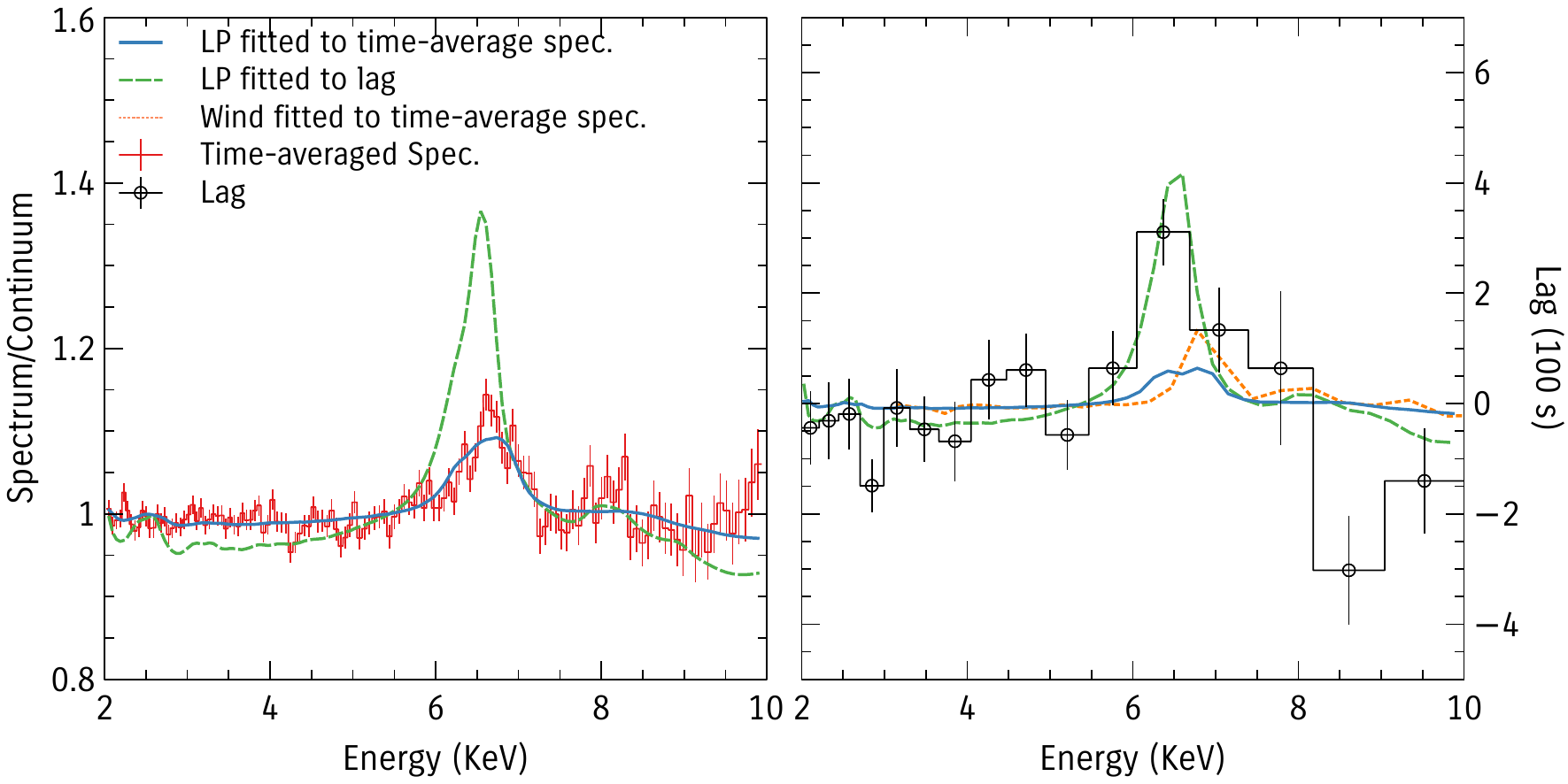}
\caption{Comparing the spectral and lag lamp-post models. {\it Left:} Ratio of the spectrum to the continuum showing the 6.7 keV feature (the spectrum from observation 7 in Table \ref{tab:obs_log} as an example). {\it Right:}  The lag spectrum from the high flux subset. The spectral and lag models refer to the lamp-post model inferred from the spectral and lag data respectively. The dotted orange line shows the lags from the wind model fitted to the time-averaged spectra shown in Figure \ref{fig:monaco}-left. The figure shows that the reflection fraction inferred from the time-averaged spectrum is not large enough to explain the lags.
\label{fig:spec_vs_lag}}
\end{figure*}

We start by fixing the model parameters to those found in the spectral modeling in section \ref{sec:rel_model}. The only free parameter is the black hole mass ($M_8$ in units of $10^8 M_\odot$), which converts the scale from $r_g$ to physical units. The black hole mass in NGC 5506 is uncertain, with stellar dispersion measurements suggesting $M_8 = 0.88$ \citep{2004MNRAS.348..207P}, and X-ray variability time scale suggesting $M_8 = 0.03$ \citep{2009MNRAS.394.2141N}.
We model the high-frequency high-flux lag spectra, and the best fit model is shown as a solid line in the right panel of Figure \ref{fig:spec_vs_lag}. We find $\chi^2=45.6$ for 15 d.o.f, and $M_8=(6^{+52}_{-2})\times10^{-3}$, closer to the X-ray variability mass estimate. This model is rejected by the lag data with high confidence ($p=6\times10^{-5}$). Although it is not unreasonable to think that increasing $M_8$ increases the lags for the same reflection fraction, doing so also reduces the frequency over which those lags are observable, and that is what puts an upper limit to the mass. {\em Fundamentally, the reflection fraction in the model that fits the time-averaged spectra is too small to account for the lags}.

To improve the fit to the lag data, we tested allowing several parameters to change.
The most significant change were found by allowing the inner radius $R_{\rm in}$ ($\Delta\chi^2=7.9$ for 1 d.o.f) and inclination $\theta$ ($\Delta\chi^2=15.1$ for 1 d.o.f) to change. The best fit model that allow for both these parameters to change is shown as the dotted line in the right panel of Figure \ref{fig:spec_vs_lag}. The parameters of the model are: $R_{\rm in} = 4^{+22}_{-3} r_g$, $\theta=(15^{+9}_{-15})^{\circ}$ and $M_8 = 0.013\pm0.003$. This model describes the lag data very well ($\chi^2=19.9$ for 13 d.o.f,  $p=0.1$).

This model, as Figure \ref{fig:spec_vs_lag} shows, have a higher reflection fraction, so larger lags can be produced for a given black hole mass, allowing it to model the lags properly. The increase in reflection fraction however, now over-predicts the strength of the line in the time-averaged spectrum, which is better described by a lower reflection fraction (left panel of Figure \ref{fig:spec_vs_lag}).

It is also possible to model the lags with a model where the ratio of the primary to the reflected normalization are not restricted to the static lamp-post case. As expected, the fit improves significantly ($\Delta\chi^2=19$ for 1 less degree of freedom), and the primary to reflected ratio is $3.9^{+2.6}_{-0.4}\%$ (parameter {\tt Np:Nr} in {\tt kynxilrev}). In other words, a significant fraction of the photons from the illuminating source need to hit the disk than reach the observer, which could be an indication of beaming. Such model however also over-predicts the strength of reflection in the time-averaged spectrum.

In summary, the lag spectra requires the reflection fraction to be {\em larger} than that observed in the time-averaged spectrum. Because of this, we find that a static lamp-post model cannot reproduce both the lag and time-averaged spectra at the same time.

\section{Discussion}\label{sec:discuss}
\subsection{Summary of the Results}
NGC 5506 shows a strong, resolved residual feature at 6.7 keV that has been attributed to either a blend of at least two narrow emission lines from Compton-thin plasma, or to a weakly-relativistic disk line. The spectral modeling of a new observation, along with old observations, cannot distinguish between the models, primarily because of the limited energy resolution of the detectors. The data suggest that the flux from the line(s) is correlated with the primary continuum flux on time scales of years (section \ref{sec:rel_model}), indicating that the line is responding to changes in the continuum.

On time scales of $\sim10$ ks, the data allows only for an upper limit on the fractional variability of the 6.7 keV feature (section \ref{sec:timing:cov}), which is consistent with its fractional flux contribution, an indication that the feature is likely varying with the continuum at these time scales.

The lag analysis in section \ref{sec:timing:delays} shows that the 6.7 keV feature is delayed with respect to the continuum, and that the delay depends on the flux level. The simplest interpretation, again, is that the 6.7 spectral feature is due to a weakly-relativistic disk line that is delayed with respect to the continuum. 

\subsection{The Static Lamp-Post Model}
Using a disk reverberation model that assume a static lamp-post geometry (section \ref{sec:timing:lag_model}) with parameters from the time-averaged spectrum fails to fit the lag spectra, under-predicting the lags. A model with a lower disk inner radius and inclination has a higher reflection fraction, and is able to reproduce the lags, but now over-predicts the contribution of reflection to the time-averaged spectrum (Figure \ref{fig:spec_vs_lag}). In other words, the lag and energy spectra, taken together, are inconsistent with a lamp-post geometry where a static point source illuminates the accretion disk. The main reason is that the reflection fraction in the time-averaged spectra is too small to account for the lags.

The failure of the lamp-post reverberation model results is similar to several other studies that attempted to model both the spectral and timing data at the same time. This was manifested for instance as wavy residuals in the lag-frequency data \cite{2018MNRAS.480.2650C} or a 3 keV dip in the lag-energy spectra \citep{2013MNRAS.434.1129K,2016MNRAS.458..200W,2016MNRAS.460.3076C,2019MNRAS.488..324I}. We note that the 3 keV dip may not be a dip, but a consequence of the fact that the reflection fraction required to model the Fe K lags is higher than that inferred from the time-averaged spectrum (e.g Figure 10 in \citealt{2016MNRAS.460.3076C} and Figure 16 in \citealt{2019MNRAS.488..324I}), similar to what we find here in NGC 5506. We note also that the statistical significance of the 3 keV dip has not always been addressed explicitly, so its existence as a separate feature may not be significant in all cases discussed in the literature \citep[e.g.][]{2017MNRAS.471.4436W}.

The main observable discrepancy between the lag and time-averaged spectra when modeled with a static lamp-post model is the enhanced reflection fraction in the former. The case of NGC 5506 presents a clear manifestation of this discrepancy. 

It is conceivable that the lag spectra, through the frequency filtering, selects regions with enhanced reflection, whose effect is averaged out in the time-average spectra. Sheared hot spots in the disk may produce such an effect. However, these are expected to be random, transient and likely present throughout the disk, and it seem unlikely that they produce a consistent picture in which reflection is enhanced when selecting some time scale and not others.

Modeling the lag data therefore clearly requires models beyond a static lamp-post geometry. Our modeling of the NGC 5506 suggests that the reflection fraction in the lag needs to be enhanced by a factor of 6 compared to that inferred from modeling the time-averaged spectrum. 

A few modifications to the simple lamp-post geometry we used here have been considered in other works. 
\cite{2016MNRAS.460.3076C} model the additional complexity in the data by including an ionization gradient in the disk. Unlike the model we used here, where the whole disk is assumed to have the same ionization parameter, they find that if the innermost
regions are highly ionized compared to the outer regions, the reverberation signatures produced through reflection from the inner parts (iron K$\alpha$ photons redshifted to 3 keV) are less observable. The line is enhanced as it is dominated by reflection from colder reflection off the outer part. We test this model in the data by allowing the density (parameter {\tt density}) and the density profile ({\tt den\_prof}) in the model to change. These provide no significant improvement in the fit ($\Delta\chi^2=0.5$ for two degrees of freedom). This is likely a result of the fact that the inner radius of the disk in NGC 5506 is relatively large, and the impassivity index is flatter ($q=3$) than the models in \cite{2016MNRAS.460.3076C}, so the ionization gradient does not provide enough contrast between the inner and outer disk provide the necessary reflection fraction enhancement.

\cite{2016MNRAS.458..200W} considers a more complicated case of a combination of hard lags produced by propagation through an extended emitting region and reverberation from a compact, point-like corona. If luminosity fluctuations propagate upwards through a collimated, vertically extended corona, it is found that the interplay between light-travel delays and propagation delays can produce lag-energy spectra where photons at 3 keV are the earliest to arrive, producing a dip at 3 keV and an enhanced line feature. A slightly different, but comparable, geometry was considered by \cite{2017MNRAS.465.3965C}, where two axial point sources illuminating an accretion disk to model the data of PG 1244+026 are used. The lags are produced from the light-travel delays between the two points and the disk and also from the assumption of different source responses for two X-ray sources. Unlike other source, strong relativistic effects are not present in NGC~5506, implying that the discrepancy between the lag, if attributed to relativistic reverberation, and the time-averaged spectra is not an effect of general relativity, but rather likely a geometrical effect.


It is worth noting that complexity beyond the lamp-post may well be required given the turbulent flow in accretion disks around black holes. The simplicity and predictive power in relativistic models is however lost. Simple fits to the data to extract important parameters such as the black hole spin is no longer a simple task.

\begin{figure}
\centering\includegraphics[width=\columnwidth]{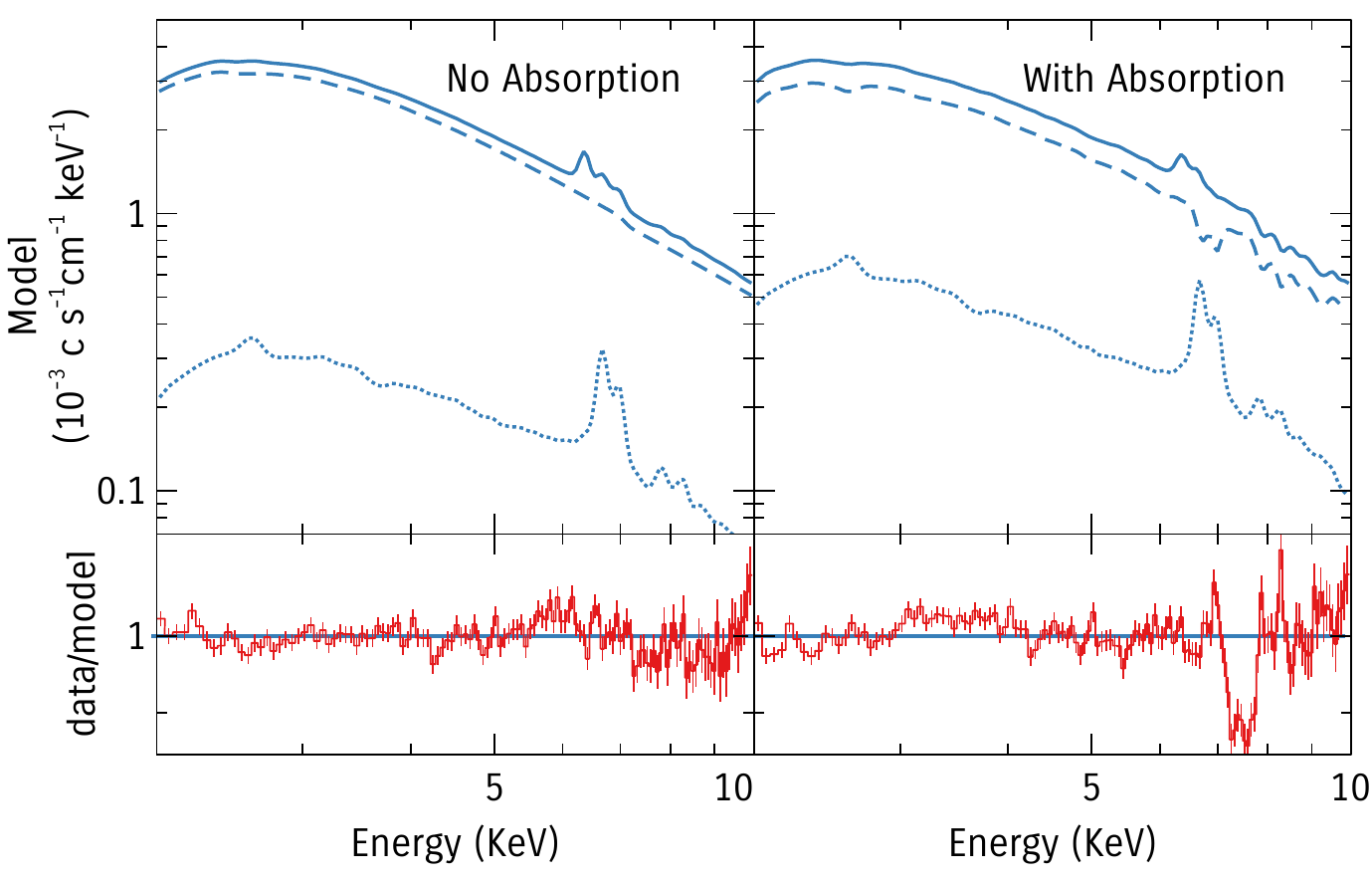}
\caption{Results of fitting the scattering model from \cite{2018MNRAS.478..971M} to the spectral data of NGC 5506. The left panel shows the spectral model when the wind is not in our line of sight. Only scattering is included. This gives a reasonable description for the spectral data, but the predicted lag is not large enough. The right panels shows the spectral model when the line of sight absorption is also included. The reflection fraction is enhanced, giving larger lags, but the model now predicts spectral features that are inconsistent with the data.
\label{fig:monaco}}
\end{figure}

\subsection{Wind Reverberation}
Given these implications for the static lamp-post relativistic model, we also consider alternative models to explain the spectral and timing observations in NGC 5506. While \cite{2010MNRAS.408.1928M, 2010MNRAS.403..196M} and \cite{2017MNRAS.467.3924T} showed that lags from a cloud at $\sim100$ \rg could explain the delays observed below 1 keV in several sources. \cite{2018MNRAS.478..971M, 2019MNRAS.482.5316M} showed that lags in the \FeK band, similar to those observed, can be produced by a distant scattering wind that is out-flowing at $\sim0.1c$. The short lags compared to the light-travel time are explained by dilution effects, where the majority of the photons in the \FeK band are primary photons with zero delay, whereas the time-delayed reprocessed photons only make a subtle contribution.

To check if this model can explain the both the spectra and the lags in NGC 5506, the observations are compared to spectral and timing products generated using the model in \cite{2018MNRAS.478..971M}. This model is not designed to fit the data, so we focus here on products that are generated only at few parameters that were found to roughly match the shape of the spectrum and lag scale. We find that the wind needs to be ionized to explain the 6.7 keV features. We also find that no simple model can model {\it both} the spectrum and the lag at the same time for the same reason the lamp-post model fails. The reflection fraction in from the model that fits the time-average data is not large enough to explain the lags (see the resulting spectrum in Figure \ref{fig:monaco}-left, and the lag in the orange dotted line in the right panel of Figure \ref{fig:spec_vs_lag}). If the lags are made larger by increasing the size of the scattering clouds or by increasing the scattering fraction outside the line of sight, the frequencies at which those lags are produced decreases compared to the observations.

If absorption in the line of sight from these same scattering clouds is also included in the model, we find that a larger lag for the same reflection fraction in the time-average spectra can be produced. This is a result of the reduced dilution in the primary continuum as the flux at the line energies is reduced due to the presence of absorption lines. This model however also predicts that several absorption lines other than those at 6.7 and 6.9 should be observed, and that means the model fails to reproduce the spectral data correctly above 7 keV as Figure \ref{fig:monaco}-right shows. The spectral data themselves show no evidence for absorption lines as the model predict.

\acknowledgments
AZ acknowledges funding under NASA grant number NNX16AH19G, through the XMM-Newton Guest Observer Facility. MM is financially supported by JSPS overseas research fellowship.

\bibliography{reference}


\end{document}